\documentclass[pdflatex, sn-standardnature]{sn-jnl}

\usepackage[separate-uncertainty=true, multi-part-units  = single]{siunitx}
\DeclareSIUnit{\sccm}{sccm}
\usepackage{braket}
\usepackage{newfloat}
\DeclareFloatingEnvironment[name={Extended Data Figure}]{suppfigure}

\begin{document}

\title[Frontier orbitals control dynamical disorder in molecular semiconductors]{Frontier orbitals control dynamical disorder in molecular semiconductors}

\author*[1]{\fnm{Alexander} \sur{Neef}}\email{alexander.neef1@gmail.com}

\author[2]{\fnm{Sebastian} \sur{Hammer}}

\author[3,4]{\fnm{Yuxuan} \sur{Yao}}

\author[5]{\fnm{Shubham} \sur{Sharma}}

\author[6]{\fnm{Samuel} \sur{Beaulieu}}

\author[7]{\fnm{Shuo} \sur{Dong}}

\author[1,8]{\fnm{Tommaso} \sur{Pincelli}}

\author[2]{\fnm{Maximillian} \sur{Frank}}

\author[1]{\fnm{Martin} \sur{Wolf}}

\author[5]{\fnm{Mariana} \sur{Rossi}}

\author[4]{\fnm{Harald} \sur{Oberhofer}}

\author[1]{\fnm{Laurenz} \sur{Rettig}}

\author[2,9]{\fnm{Jens} \sur{Pflaum}}

\author*[1,8]{\fnm{Ralph} \sur{Ernstorfer}}\email{ernstorfer@tu-berlin.de}

\affil[1]{\orgdiv{Department of Physical Chemistry}, \orgname{Fritz Haber Institute of the Max Planck Society}, \orgaddress{\street{Faradayweg 4-6}, \city{Berlin}, \postcode{14195}, \country{Germany}}}

\affil[2]{\orgdiv{Experimental Physics VI}, \orgname{University of Wuerzburg}, \orgaddress{\street{Am Hubland}, \city{Wuerzburg}, \postcode{97074}, \country{Germany}}}

\affil[3]{\orgdiv{Department of Chemistry, TUM School of Natural Sciences},\orgname{Technical University Munich}, \orgaddress{\street{Lichtenbergstr. 4},\postcode{85748}, \city{Garching b. München}, \country{Germany}}}

\affil[4]{\orgdiv{Chair for Theoretical Physics VII and Bavarian Center for Battery Technology},\orgname{University of Bayreuth}, \orgaddress{\street{Universitätsstr. 30},\postcode{95447}, \city{Bayreuth}, \country{Germany}}}

\affil[5]{\orgname{Max Planck Institute for the Structure and Dynamics of Matter}, {\postcode{22761}, \city{Hamburg}, 
 \country{Germany}}}

\affil[6]{\orgdiv{CELIA}, \orgname{Université de Bordeaux–CNRS–CEA}, \orgaddress{\city{Bordeaux}, \postcode{UMR5107}, \country{France}}}

\affil[7]{\orgdiv{Beijing National Laboratory for Condensed Matter Physics}, \orgname{Institute of Physics, Chinese Academy of Sciences}, \orgaddress{\city{Bejing}, \postcode{100190}, \country{China}}}

\affil[8]{\orgdiv{Institut für Optik und Atomare Physik}, \orgname{Technical University Berlin}, \orgaddress{\street{Straße des 17. Juni 135}, \city{Berlin}, \postcode{10623}, \country{Germany}}}

\affil[9]{\orgname{Center for Applied Energy Research e.V.}, \orgaddress{\street{Magdalene-Schoch-Straße 3}, \city{Wuerzburg}, \postcode{97074}, \country{Germany}}}

\abstract{Charge transport in organic semiconductors is limited by dynamical disorder. Design rules for new high-mobility materials have therefore focused on limiting its two foundations: structural fluctuations and the transfer integral gradient. However, it has remained unclear how these goals should be translated into molecular structures. Here we show that a specific shape of the frontier orbital, with a lack of nodes along the long molecular axis, reduces the transfer integral gradient and therefore the dynamical disorder. We investigated single crystals of the prototypical molecular semiconductors pentacene and picene by angle-resolved photoemission spectroscopy and dynamical disorder calculations. We found that picene exhibits a remarkably low dynamical disorder. By separating in- and out-of-plane components of dynamical disorder, we identify the reason as a reduced out-of-plane disorder from a small transfer integral derivative. Our results demonstrate that molecules with an armchair $\pi$-electron topology and same-phase frontier orbitals like picene are promising molecular building blocks for the next generation of organic semiconductors.}

\maketitle


\section*{1.Introduction}
Since the dawn of organic electronics, its promise has been to achieve device functionality with flexibility unattainable with conventional semiconductors \cite{Forrest2004Apr}. For lighting applications, OLEDs have fulfilled this promise. Organic photovoltaic cells have started to catch up with silicon cells with the advent of non-fullerene acceptors \cite{Lin2015Feb}. Organic transistors, on the other hand, are lagging behind \cite{Schweicher2020Mar}. Some fundamental questions related to maximizing the mobility of organic semiconductors at room temperature and the feasibility of band transport in thin films and single crystals remain unanswered. Understanding the underlying charge transport mechanisms is essential to approach these questions.

Here we focus on single crystals of molecular semiconductors as well-defined reference systems. A key ingredient in the description of charge transport in molecular semiconductors is dynamical disorder \cite{Troisi2006Mar}. Dynamical disorder arises from the coupling of structural fluctuations of the molecules inherent to soft and flexible organic materials to the intermolecular transfer integrals. The ensuing distributions of transfer integrals have widths $\sigma_t$ (the standard deviation) that are of similar magnitude as the mean value of transfer integrals $t$ at room temperature, i.e. $\sigma_t/t\gtrsim0.3$ \cite{Fratini2020May}.

This substantial disorder leads to Anderson localization of the electronic states. The localization effects are particularly prominent at the tails of the electronic bands where the carriers relevant for charge transport reside. Due to the fluctuating nature of the disorder, the states continuously change, they localize and delocalize depending on the transient transfer integral landscape.

Fratini \textit{et al.} have incorporated these localization phenomena into their transient localization theory (TLT) \cite{Fratini2016Apr} which has been able to predict the charge mobility of various molecular semiconductors reliably. In TLT, the crucial quantity determining the mobility $\mu$ is the \textit{localization length} $L$ as $\mu\propto L^2$. The localization length, a measure of the spatial extent of a state, is determined by two factors: (1) dynamical disorder and (2) the resilience of electronic interaction to dynamical disorder.

As dynamical disorder governs the localization length and hence the mobility, substantial efforts have been invested into reducing $\sigma_t$. One approach is to reduce the amplitude of structural fluctuations $\sigma_r$. This approach is promising as the range of $\sigma_r\sim0.1$ to $\SI{0.3}{\angstrom}$ is quite significant. A reduction of $\sigma_r$ has indeed been achieved by strain \cite{Kubo2016Apr} or by adding side groups \cite{Illig2016Feb}, in the latter case, however, with an accompanying undesirable change of the electronic structure. The approach we pursue here is a different one. We aim to reduce the coupling between structural fluctuations and intermolecular transfer integrals.

\begin{figure}
    \centering
    \includegraphics{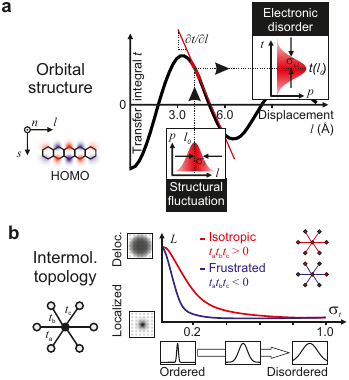}
    \caption{\textbf{\textbar~Electronic disorder and localization in molecular semiconductors. a}, Schematic dependence of the transfer integral $t$ on the long axis displacement $l$ for the HOMO of a face-on pentacene dimer. The molecular structure of pentacene, its long ($l$), short ($s$), and normal ($n$) axes, and the HOMO are indicated on the left. Thermal fluctuations of $l$ with standard deviation $\sigma_{l}$ lead to fluctuations of the transfer integral with the standard deviation $\sigma_{t, l}=\partial t/\partial l\times\sigma_l$. The standard deviation of the electronic disorder is proportional to the derivative of the transfer integral at the equilibrium position $l_0$. \textbf{b}, Schematic dependence of the localization length of band tail states on the width of the electronic disorder for an isotropic and frustrated intermolecular topology. The relationship is modeled for a 2D tight-binding model with three distinct nearest-neighbor transfer integrals: $t_a$, $t_b$, and $t_c$. For the isotropic and frustrated topologies, the structure of the highest-lying states is shown.}
    \label{F1}
\end{figure}

The dependence of the transfer integral on an intermolecular degree of freedom (DOF) $r$ transforms structural fluctuations into electronic disorder. In Fig.~\ref{F1}a, the dependence of the highest occupied molecular orbital (HOMO) transfer integral on the long-axis translation $l$ in a pentacene dimer is shown. The electronic disorder from this DOF is determined by the magnitude of the translational disorder $\sigma_l$ and the derivative $\partial t/\partial l$ at the position $l_0$. While the position $l_0$ is determined by the crystal structure, the shape of $t(l)$ is related to the structure of the underlying orbital. We explore here how the structure of the frontier orbital yields a desirable, i.e. slowly varying, $t(l)$-curve.

Molecular semiconductors commonly form layered crystal structures with a clearly defined high-mobility plane spanned by the lattice vectors $\mathbf{a}$ and $\mathbf{b}$ and weak interactions between the layers. Their electronic structure can hence be described by a 2D tight-binding (TB) model with six nearest neighbors and three distinct transfer integrals ($t_a,t_b,t_c$, see Fig.~\ref{F1}b and Methods) \cite{Fratini2017Oct}. The band structure and the localization properties crucially depend on the sign combination of the three transfer integrals. If the product of the transfer integrals is positive ($t_at_bt_c>0$), the transfer integrals are called \textit{isotropic}. We introduce the term \textit{frustrated} transfer integrals to denote a negative product ($t_at_bt_c<0$), in analogy to geometric frustration of the antiferromagnetic state on a triangular lattice \cite{Wannier1950Jul}.

By adding electronic disorder to the model, quantified by the standard deviation of the transfer integral distribution $\sigma_t$, the different localization properties of isotropic and frustrated transfer integrals become apparent. Fig.~\ref{F1}b shows the localization length of tail states as a function of $\sigma_t$. The tail states formed by isotropic transfer integrals resist localization much longer than those formed by frustrated transfer integrals. They are therefore more resilient to disorder.

In this work, we investigate these two cases based on the prototypical molecular semiconductors pentacene and picene by combining experimental and computational studies of their electronic structure and dynamical disorder. Although pentacene and picene both form a herringbone packing, their valence bands differ substantially. Their dynamical disorder is examined by a combination of molecular dynamics (MD) simulations and fragment-orbital density functional theory (FO-DFT). We show that the dynamical disorder is reduced in picene compared to pentacene. The reason is the favorable structure of the picene HOMO. The structure of the HOMO also leads to isotropic transfer integrals and hence enhanced resilience to disorder.

\section*{2.Results}
We performed angle-resolved photoemission spectroscopy (ARPES) to obtain the experimental valence band structures of pentacene and picene single crystals. The energy resolution of the experiment was $E_\text{FWHM}=\SI{150}{\milli\electronvolt}$. For details of the setup, the reader is referred to \cite{Maklar2020Dec}.

\begin{figure}
    \centering
    \includegraphics{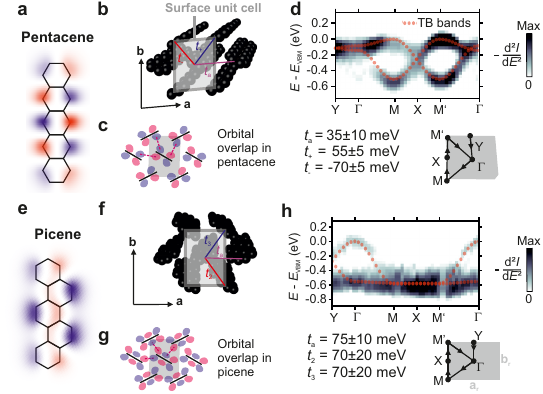}
    \caption{\textbf{\textbar~Experimental valence band structure of pentacene and picene. a}, The HOMO of pentacene changes sign along the long axis. \textbf{b}, The crystal structure of pentacene and \textbf{c}, the corresponding orbital overlap of HOMOs. \textbf{d}, Valence band structure of pentacene obtained by ARPES to which TB bands were fitted, resulting in the shown transfer integrals. The data shows the second derivative of the photoemission intensity with respect to energy, $\frac{\text{d}^2I}{\text{d}E^2}$. The data is extracted along the given path in reciprocal space. \textbf{e}, The HOMO of picene features the same sign along the long axis. \textbf{f-h}, Crystal structure, orbital overlap, and experimental band structure for picene.}
    \label{F2}
\end{figure}

The valence band of pentacene is branched into a lower and an upper part due to the presence of two inequivalent molecules in the triclinic unit cell (see Fig.~\ref{F2}a-d). By fitting the TB model to the bands, we obtain the transfer integrals $t_a=\SI{35(10)}{\milli\electronvolt}$ connecting dimers separated by $\mathbf{a}$, $t_+=\SI{55(5)}{\milli\electronvolt}$, connecting dimers separated by $\frac{\mathbf{a}+\mathbf{b}}{2}$, and $t_-=\SI{-70(5)}{\milli\electronvolt}$ connecting dimers separated by $\frac{\mathbf{a}-\mathbf{b}}{2}$. These compare well with previous results, e.g. ($\SI{18}{\milli\electronvolt}$, $\SI{47}{\milli\electronvolt}$, $\SI{-56}{\milli\electronvolt}$) obtained by ARPES \cite{Hatch2009Aug} and ($\SI{24}{\milli\electronvolt}$, $\SI{43}{\milli\electronvolt}$, $\SI{-79}{\milli\electronvolt}$) obtained by computations \cite{Yoshida2008Jun}. Note that the orbital overlap in pentacene leads to the measured frustrated transfer integrals, i.e. $t_at_+t_-<1$.

Picene crystallizes in a herringbone structure similar to pentacene in a more symmetric monoclinic unit cell. The individual molecules have a lower symmetry ($C_{2v}$) than pentacene ($D_{2h}$) and lack an inversion center. The unit cell correspondingly lacks inversion symmetry. Taking these symmetries into account, the three distinct transfer integrals $t_a$, $t_2$, and $t_3$ connect different dimers in picene. The latter two quantify the interaction between picene dimers where the two-ring (three-ring) edges of picene are closest (see Fig.~\ref{F2}e-h). Despite the similar crystal packing, the picene valence band exhibits a markedly different structure than that of pentacene. By fitting the TB model (see Methods for details), we extracted the transfer integrals $t_a=\SI{75(10)}{\milli\electronvolt}$, $t_2=\SI{70(20)}{\milli\electronvolt}$, and $t_3=\SI{70(20)}{\milli\electronvolt}$. These agree with computational results, e.g. ($\SI{70}{\milli\electronvolt}$, $\SI{66}{\milli\electronvolt}$, $\SI{54}{\milli\electronvolt}$) \cite{Nguyen2015May}. Only one experimental study has reported the band structure with a lower $t_a=\SI{45}{\milli\electronvolt}$ and without reporting the other two transfer integrals \cite{Xin2012May}. Due to the different structure of the picene HOMO along the short molecular axis, the orbital overlap leads to isotropic transfer integrals in picene.

Having extracted the transfer integrals by ARPES, we now turn to the analysis of the structural fluctuations. For that purpose, we conducted \textit{ab initio} quality MD simulations at room temperature and extracted the relative displacements of nearest neighbors based on their local coordinate systems, yielding six DOF (three translations and three rotations). On average, the translations are larger in picene ($\sigma_r=\SI{0.30}{\angstrom}$) than in pentacene ($\sigma_r=\SI{0.21}{\angstrom}$, see Ext.~Data~Fig.~\ref{SuppFig:sigma_r}).

To obtain the dynamical disorder, we extracted next-neighbor dimers from MD and computed the HOMO transfer integrals by FO-DFT. The mean values of the transfer integrals for pentacene ($\SI{32.0}{\milli\electronvolt}$, $\SI{39.5}{\milli\electronvolt}$, $\SI{-78.8}{\milli\electronvolt}$) and picene ($\SI{64.4}{\milli\electronvolt}$, $\SI{63.3}{\milli\electronvolt}$, $\SI{60.3}{\milli\electronvolt}$) compare well with the values of the TB fit.

We quantified the amplitude of dynamical disorder by the standard deviation $\sigma_t$ of the distribution of transfer integrals (see Ext.~data.~Fig.~\ref{SuppFig:t_statistics} for the full distribution). In pentacene ($\sigma_{t, a} = \SI{12.0}{\milli\electronvolt}$, $\sigma_{t, +}=\SI{18.0}{\milli\electronvolt}$, $\sigma_{t, -}=\SI{18.4}{\milli\electronvolt}$), the dynamical disorder is significantly larger than in picene ($\SI{10.1}{\milli\electronvolt}$, $\SI{13.8}{\milli\electronvolt}$, $\SI{15.3}{\milli\electronvolt}$). The results for the dynamical disorder in pentacene are in good agreement with previous computations ($\SI{11.4}{\milli\electronvolt}$, $\SI{18.2}{\milli\electronvolt}$, $\SI{16.4}{\milli\electronvolt}$) \cite{Fratini2017Oct}.

The parameter determining the localization properties and hence the mobility is the relative dynamical disorder $\sigma_t/t$ which we show in Fig.~\ref{F3}a. For all dimers, picene features a smaller relative dynamical disorder than pentacene. Furthermore, the overall relative disorder, defined as $\sigma_\text{tot}=\sqrt{\sigma_a^2+\sigma_+^2+\sigma_-^2}/\sqrt{t_a^2+t_+^2+t_-^2}$, is exceptionally low in picene ($\sigma_\text{tot}=0.185$) compared to other known compounds, e.g. $\sigma_\text{tot}=0.295$ in rubrene \cite{Fratini2020May}, which is the benchmark material for high mobilities in molecular semiconductors.

Why is the dynamical disorder so low in picene despite the large structural fluctuations? To explore this question we separate the DOF into out-of-plane and in-plane components, where the plane is spanned by the short and normal axes. The translation along the long molecular axis and the rotations around the short and normal axes have out-of-plane components, whereas translations along the short and normal axes and the rotation around the long axis are entirely in-plane. We found no correlation between in-plane and out-of-plane DOF and between the individual out-of-plane DOF. The individual in-plane DOFs are, however, strongly correlated (see Ext.~data.~Fig.~\ref{SuppFig:MDcorrelation}).

We can then write the overall absolute disorder as
\begin{equation*}
    \sigma_t = \sqrt{\sigma_\text{out}^2+\sigma_\text{in}^2} =\sqrt{\sigma^2_{t, l} + \sigma_{t, \phi_s}^2 + \sigma_{t, \phi_n}^2 + \sigma^2_{t, (s, n, \phi_l)}},
\end{equation*}
where the first three terms on the right side encompass the out-of-plane and the last term the in-plane disorder. Each out-of-plane terms is given by the linear approximation $\sigma_{t, r}=\partial t/\partial r\times\sigma_r$. The relative disorder is obtained by normalizing the absolute disorder with the transfer integral obtained by ARPES. Finally, the overall relative disorder and its in- and out-of-plane components are visualized as a vector. This procedure is shown for the $t_a$ integral in pentacene in Fig.~\ref{F3}b and reveals a large out-of-plane component of the dynamical disorder.

In Fig.~\ref{F3}c, the vector representation is shown for all transfer integrals in the two materials. The vectors fall into two categories. In the first category (encompassing $t_-$ in pentacene and all transfer integrals in picene), the relative dynamical disorder is dominated by the in-plane component. In the second category (including $t_a$ and $t_+$ in pentacene), the out-of-plane component dominates or plays a significant role.

Why, then, is the out-of-plane component so low for the pentacene $\frac{\mathbf{a}-\mathbf{b}}{2}$-dimer and all picene dimers? In the pentacene dimer, the molecules are only marginally displaced along the long axis ($l_0\approx0$), such that it is at a sweet spot (i.e. an extremal point) of the transfer integral curve (compare Fig.~\ref{F1}a). Therefore, the transfer integrals only change little by out-of-plane fluctuations and the dynamical disorder is correspondingly low. The $\mathbf{a}$- and $\frac{\mathbf{a}+\mathbf{b}}{2}$-dimers, on the other hand, are displaced by $l_0\approx\SI{2}{\angstrom}$ and the derivative $\partial t/\partial l$ and the out-of-plane disorder are hence large. In picene, all dimers feature low out-of-plane disorder and the reason is different than in pentacene. This can be understood by replacing the pentacene HOMO in Fig.~\ref{F1}a with the picene HOMO (Fig.~\ref{F2}e). The lack of nodal planes of the picene HOMO along the long axis leads to exceptionally small variations of $t$ with out-of-plane fluctuations.
With the computed dynamical disorder and the experimental transfer integrals we then conducted simulations to obtain information about the localization properties. We used the TB model described above and drew the transfer integrals from the distribution $P(t)\propto\exp{-(t-t_0)^2/(2\sigma_t^2)}$, where $t_0$ is the mean value of a given transfer integral. The resulting density of states (DOS) is shown in Fig.~\ref{F3}d. For pentacene, the DOS is the largest close to the valence band maximum (VBM) because its frustrated transfer integrals lead to flat bands at that energy. This is reversed for picene which features a low DOS at the VBM. At energies significantly larger than $E_\text{VBM}$, we find tail states in pentacene and to a much lesser extent in picene. Note that the energy resolution of the ARPES experiment ($E_\text{FWHM}=\SI{150}{\milli\electronvolt}$) is insufficient to resolve these tail states.

To analyze the localization properties, we calculated the mean inverse participation ratio (IPR), a common measure of delocalization \cite{Thouless1974Oct} (see Fig.~\ref{F3}e). The states in the band center of both materials are highly delocalized, whereas the tail states above the VBM are much more localized. The tail states in pentacene extend to $E-E_\text{VBM}=\SI{50}{\milli\electronvolt}$ and have an $\text{IPR}\approx10$. Picene, by contrast, features tail states that are much more delocalized with an $\text{IPR}\approx100$, one order of magnitude larger.

\section*{3.Discussion}
\begin{figure}
    \centering
    \includegraphics{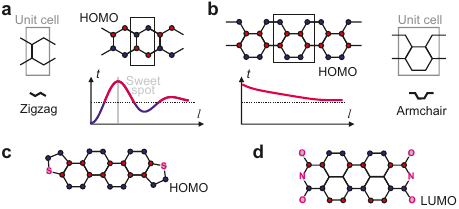}
    \caption{\textbf{\textbar~Frontier orbital design. a}, The HOMO of graphene nanoribbons with a zigzag edge and the corresponding dependence of the intermolecular transfer integral. \textbf{b}, Same as \textbf{a} for graphene nanoribbons with an armchair edge. \textbf{c}, HOMO of ChDT (see main text for full name), a sulfur-based molecular semiconductor showcasing a picene-like HOMO. \textbf{d}, LUMO of perylene diimide, a molecular building block of benchmark materials with high electron mobilities.}
    \label{F4}
\end{figure}

The essential difference between pentacene and picene is the HOMO structure. The lack of nodes along the long axis reduces dynamical disorder in picene. How can such orbital structures be rationally designed? To understand the relation between $\pi$-topology, i.e. the structure of the connected network of $sp^2$-hybridized atoms, and orbital structure better, we borrow from the theory of graphene nanoribbons (GNRs) \cite{Nakada1996Dec}. GNRs are networks of sp$^2$-hybridised carbon atoms periodic along one dimension. They can be distinguished by their dominant edge shape which is either of zigzag (Fig.~\ref{F4}a) or armchair type (Fig.~\ref{F4}b). The zigzag GNRs (zGNRs) are akin to the structure of pentacene. Their VBM is at the edge of the Brillouin zone. In other words, their frontier orbitals change sign with each unit cell (or every two carbon atoms). This is different for armchair GNRs (aGNRs), which are similar to picene. Their VBM is at the center of the Brillouin zone and, correspondingly, the sign of the frontier orbitals remains the same on each unit cell. If the orbital within the unit cell does not change along the periodic direction (i.e. the long axis), this leads to a frontier orbital with no nodes along the long axis. The intermolecular transfer integrals oscillate for the frontier orbitals of zGNRs but only slowly change for those of aGNRs.

Frontier orbitals with no nodes along the long axis also occur in sulfur-based molecular semiconductors such as chryseno[2,1-b:8,7-b']dithiophene (ChDT) \cite{Yamamoto2018Jan} (see Fig.~\ref{F4}c) or perylene diimide (PDI) that are isoelectronic to aGNRs. Several recent studies have played with the idea of designing picene-like frontier orbitals \cite{Yamamoto2018Jan, Mitani2022Nov, Banks2023Sep}. Yamamoto \textit{et al.} have reported high mobilities for ChDT-based transistors, thus proving the potential of frontier orbital design. A further example is the material diphenanthro[1,2-b:2',1'-d]thiophene (DPT) for which a patented transistor has been recently developed \cite{Okamoto2019Aug}. For electron-transporting materials, PDI-based materials have shown the highest mobilities \cite{Molinari2009Feb, Okamoto2020May}. The situation is slightly different for these because they exhibit brickwall-type packings, i.e. with only four interacting nearest neighbors and different structural fluctuations. Future studies should project the dynamical disorder of electron-transporting materials to in-plane/out-of-plane components and connect the results to the structure of the frontier orbital.

We propose a new design strategy for future high-performance MSCs. It starts by optimizing the $\pi$-topology of the core structure to get the desired frontier orbital shape with no nodes along the long molecular axis. In the second step, the core can be appended by side groups to optimize the crystal structure and ensure solution processability.

We point out that reduced dynamical disorder is only one parameter that needs to be optimized. Getting rid of defects, impurities, and the Schottky barriers that depend on the valence band mismatch and determining the injection of charge carriers at the interfaces is just as important. The lack of control over the structure and purity of picene-based transistors might explain why it has not consistently shown high mobilities. Furthermore, we have only analyzed herringbone packings here. The concepts can, however, be transferred to other packing motifs. In most brickwall and slipped $\pi$-stack packings (e.g. TIPS-pentacene and rubrene), the long molecular axis lies in-plane. Same-phase frontier orbitals then reduce in- and not out-of-plane dynamical disorder.

\section*{4.Conclusion}
In conclusion, by combining angle-resolved photoemission spectroscopy, molecular dynamics, and fragment-orbital DFT, we analyzed the dynamical disorder and localization effects in the paradigmatic organic semiconductors picene and pentacene. While the dynamical disorder stemming from in-plane degrees of freedom is similar in both materials, the contribution from out-of-plane motions is substantially reduced in picene. Crucially, this also reduces the overall dynamical disorder. The underlying reason is the beneficial structure of the picene HOMO with a lack of nodes along the long molecular axis.

Molecules with armchair edges generally feature frontier orbitals with a lack of nodes along the long molecular axis. We thus envision them and molecules with similar $\pi$-topologies to play a crucial role in reducing dynamical disorder in the next generation of molecular semiconductors.

Here we complemented an experimental determination of the band structure with computational techniques to obtain the dynamical disorder and localization properties in molecular semiconductors. Future studies should provide experimental validations of the latter quantities. The localization length of tail states is experimentally accessible by angle-resolved photoemission spectroscopy with an energy resolution $<\SI{50}{\milli\electronvolt}$ and temperature-dependent measurements could directly reveal disorder-induced localization in molecular semiconductors. This would provide the missing observational link between molecular/crystal structure and charge mobility.

\textbf{Data availability.} The data used in this study is available in this repository: https://doi.org/10.5281/zenodo.10228043.

\textbf{Conflicts of interest.} The authors declare no conflicts of interest.

\newpage

\begin{figure}
    \centering    \includegraphics[width=\textwidth,height=\textheight,keepaspectratio]{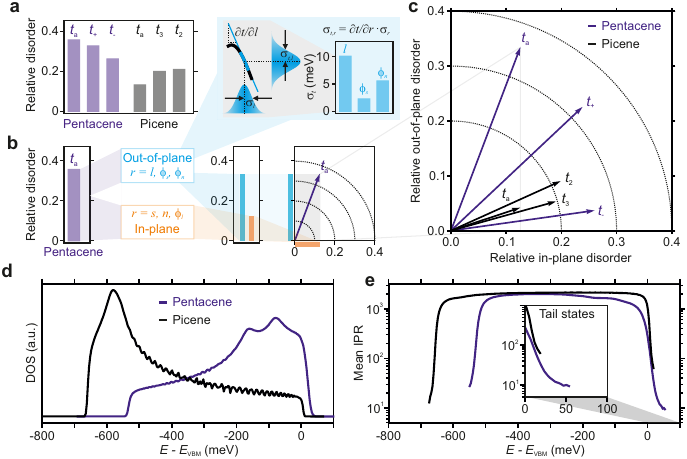}
    \caption{\textbf{\textbar~Dynamical disorder and localization properties. a}, 
    Relative dynamical disorder obtained via MD+FO-DFT for the three dimer pairs in pentacene and picene single crystals. \textbf{b}, Projection of the dynamical disorder on out-of- and in-plane degrees of freedom for the $\mathbf{a}$-dimer in pentacene. The out-of-plane dynamical disorder consists of three independent terms each obtained as $\sigma_{t, r}=\partial t/\partial r \sigma_r$. These are then normalized by the transfer integral $t_a=\SI{35}{\milli\electronvolt}$. The independent in-plane dynamical disorder is then $\sigma_\text{in}=\sqrt{\sigma^2-\sigma_\text{out}^2}$. The resulting terms are visualized as a vector. \textbf{c}, Vectorial representation of the relative dynamical disorder for all transfer integrals. \textbf{d}, Density-of-states obtained from a disorder simulation with the transfer integrals from ARPES and the electronic disorder from the simulations. The energy is referenced to the position of the VBM with no disorder. \textbf{e}, The inverse participation ratio, a measure of the delocalization of states, for pentacene and picene. Note the logarithmic y-axis. The inset shows the localization properties of the tail states at the VBM relevant for charge transport}
    \label{F3}
\end{figure}

\clearpage
\newpage

\section*{5.Methods}
\paragraph{Angle-resolved photoemission spectroscopy}
We used a homebuilt high-harmonic generation (HHG) setup operating at $\SI{500}{\kilo\hertz}$ to provide XUV pulses for photoemission~\cite{Puppin2019Feb} with the following parameters: Photon flux: up to $\SI{2e11}{photons\per\second}$ on target, Spectral width: $\SI{150}{\milli\electronvolt}$ FWHM, Photon energy: $\SI{21.7}{\electronvolt}$, Pulse duration: 35 fs.
To induce photoconductivity and suppress sample charging~\cite{Machida2010Apr}, we illuminated the MSC crystals with photons in the visible spectrum ($\SI{3.0}{\electronvolt}$ for picene, $\SI{1.8}{\electronvolt}$ for pentacene). The XUV pulses were incident at $\ang{65}$ with respect to the surface normal and p-polarized. The MSC crystals were glued with conductive UHV glue to a copper sample holder and then mounted on a 6-axis cryogenic manipulator (SPECS GmbH) and cleaved at a base pressure of \SI{8e-11}{\milli\bar}. The data was acquired at room temperature using a time-of-flight momentum microscope (METIS1000, SPECS GmbH), allowing the detection of each photoelectron as a single event. The resulting 3D photoemission intensity data is hence a function of the two components of the parallel momentum and the electron kinetic energy: $I(k_x,k_y,E)$. The overall energy and momentum resolutions of the experiment are $\Delta E=\SI{150}{\milli\electronvolt}$ and $\Delta k=\SI{0.08}{\per\angstrom}$, respectively~\cite{Maklar2020Dec}. Due to the damage that the XUV radiation causes in the MSCs, we maximized the field of view (field aperture diameter: $\SI{500}{\micro\meter}$) and therefore the number of probed molecules, while still maintaining the momentum resolution at the relevant kinetic energies.

\paragraph{Crystal growth} All crystals were grown by means of horizontal physical vapor deposition~\cite{Kloc2010, Laudise1998May}. 
Pentacene crystals were grown from starting material purified by two-fold horizontal gradient sublimation as described in~\cite{Hammer2020Nov}.
For Picene, crystals were grown at approximately \SI{240}{\degreeCelsius} over a period of \SI{150}{\hour} with nitrogen as inert transport gas (6N purity) at a flow rate of \SI{30}{\sccm}.
To minimize thermal stress on the as-grown crystals, we applied a cooling ramp to room temperature over \qty{3}{\hour}.
For both, pentacene and picene, the procedure yielded large plate-like crystals as shown in~\ref{SuppFig:ARPES}a)

\paragraph{General tight-binding model}
Most MSCs form layered structures in the bulk and feature a plane of high mobility, with significantly smaller mobility perpendicular to it. Within this plane, spanned by the lattice vectors $\mathbf{a}$ and $\mathbf{b}$, there are two symmetry-inequivalent molecules per unit cell. Each molecule has six nearest neighbors at the positions $\pm \mathbf{a}$, $\pm(\mathbf{a}+\mathbf{b})/2$, and $\pm(\mathbf{a}-\mathbf{b})/2$. The next neighbors are coupled by three distinct transfer integrals $t_a$, $t_b$, and $t_c$. We solve this commonly used 2D tight-binding model~\cite{Fratini2017Oct} and obtain two bands per molecular orbital. The Hamiltonian of the model is given by
\begin{equation*}
    H(\mathbf{k}) = \left(\begin{array}{cc}h_0(\mathbf{k}) & h_1(\mathbf{k})\\ h_1^*(\mathbf{k}) & h_0(\mathbf{k}) \end{array}\right),
\end{equation*}
where $h_0(\mathbf{k}) = 2t_a\cos \mathbf{k}\cdot\mathbf{a}$ and $h_1(\mathbf{k})$ depends on the symmetry of the molecule and the unit cell.
Pentacene with molecular $D_{2h}$ symmetry has an inversion center and features a triclinic unit cell. The off-diagonal term in the Hamiltonian is
$$
h_1(k) = 2t_+ \cos\left(\mathbf{k}\cdot\frac{\mathbf{a}+\mathbf{b}}{2}\right) + 2t_- \cos\left(\mathbf{k}\cdot\frac{\mathbf{a}-\mathbf{b}}{2}\right),
$$
where $t_b$ and $t_c$ have been replaced by $t_+$ and $t_-$, the transfer integrals between dimers connected by the vectors $\frac{\mathbf{a}+\mathbf{b}}{2}$ and $\frac{\mathbf{a}+\mathbf{b}}{2}$, respectively.
Picene with molecular $C_{2v}$ symmetry lacks an inversion center and features a monoclinic unit cell. The off-diagonal term in the Hamiltonian is
$$
h_1(k) = t_2(e^{ik\frac{\mathbf{a}+\mathbf{b}}{2}} + e^{-ik\frac{\mathbf{a}-\mathbf{b}}{2}}) + t_3(e^{-ik\frac{\mathbf{a}+\mathbf{b}}{2}} + e^{ik\frac{\mathbf{a}-\mathbf{b}}{2}}),
$$
where $t_b$ and $t_c$ have been replaced by $t_2$ and $t_3$, the transfer integrals of picene dimers meeting at a 2(3)-ringed edge, respectively.

\paragraph{Molecular dynamics simulations}
The molecular dynamics simulations of picene and pentacene were performed with a machine-learned interatomic potential (MLIP) trained on DFT data. We employed the MACE equivariant message passing architecture~\cite{Batatia2022mace}. To train the MLIP, an initial pool of structures for supercell size of $2 \times 2\times 1$ at \SI{295}{K} was generated with the organic force-field foundation model MACE-OFF23~\cite{kovacs2023mace}. From this simulation, we selected \SI{900} diverse structures by applying a farthest point sampling algorithm~\cite{eldar1997farthest} based on local SOAP (Smooth Overlap of Atomic Positions)~\cite{bartok2013representing} descriptors of atomic coordinates. Energies and forces for subsets of these structures were computed with density-functional theory (DFT). All DFT calculations were performed with the FHI-aims~\cite{blum2009ab} program package, using the PBE exchange-correlation functional with many-body-dispersion~\cite{tkatchenko2012accurate} corrections to account for long-range van der Walls interactions.  Predictions were performed on a separate test set to validate the potential. For picene, the potential achieved an RMSE on forces of \SI{11.1}{meV/\angstrom}  and an RMSE on energies of \SI{0.1}{meV/atom}. For pentacene, the RMSE on forces was of \SI{9.0}{meV/\angstrom} and the RMSE on energies was of \SI{0.4}{meV/atom}. We performed \SI{2}{ns} long MD simulation runs for different supercell sizes ($3 \times 3 \times 2$ and $5\times 5 \times 5$) at \SI{80}{K} and \SI{295}{K} to establish the stability of the potential. Moreover, we checked that the potentials could reproduce harmonic phonon frequencies and displacements computed with the same level of DFT, obtaining a maximum deviation on frequencies of \SI{11.09}{cm^{-1}} and average deviation of \SI{1.73} {cm^{-1}}.

\paragraph{Extracting dimer degrees of freedom}
From the molecular dynamics simulation, we extracted all nearest neighbor dimers and sorted them by the type of neighbor (i.e. $a$, $(a+b)/2$, $(a-b)/2$). We projected the vectors between their centroids onto the principal axes of one molecule (i.e. long, short, and normal axis) and obtained the relative angles by aligning the molecules. This yields six intermolecular degrees of freedom (3 translations: $l, s, n$, 3 rotations: $\phi_l, \phi_s, \phi_n$). We obtained the transfer integrals by generating the dimers from these DOFs and extracting the transfer integrals via FO-DFT, as explained next. We furthermore computed the partial derivatives $\partial t/\partial r$ by extracting the transfer integrals along the trajectory $r_0 + \delta r$ and leaving the other DOFs unchanged.

\paragraph{FO-DFT calculations}
Electronic coupling elements for hole transfer were calculated with the H$^{2n}$@DA variant of the Fragment-orbital density-functional theory (FO-DFT) method ~\cite{fodft2016, fodft2017} as implemented in the FHI-aims package.~\cite{blum2009ab} Thereby, hole couplings t$_{ab}$ between a Donor a and an Acceptor b are estimated from the Kohn-Sham Matrix element between the highest occupied molecular orbitals of both fragments. Kohn-Sham wavefunctions were expanded using a “tier2” basis set and real space integration grids were set to ‘tight’ standards. The SCF cycle was considered converged as soon as changes to the electronic energy levels fell below $10^{-6}$eV. Exchange-correlation energies were approximated by the PBE functional\cite{pbe1996}. Given the fact that coupling calculations were static, and moreover only considered fragments of the dimers, van der Waals interactions were not considered at this stage. All dimers were generated by moving and rotating pentacene or picene molecules along their principal axes (including mixed displacements).

\paragraph{Tight-binding model including disorder}
A Hamiltonian was initialized with a lattice spanning 80x80 sites and periodic boundary conditions. Next neighbors are connected by transfer integrals drawn from a probability distribution. The mean value of the distribution is the transfer integral extracted via ARPES. The width of the distributions is given by the dynamical disorder obtained by MD + FO-DFT. The model was run with 500 different initializations. The inverse participation ratio (IPR) was calculated for each state as \cite{Thouless1974Oct}
\begin{equation*}
    \text{IPR} = 1/\left[\frac{\sum_i \mid c_i \mid ^4}{\left(\sum_i \mid c_i \mid ^2\right)^2}\right].
\end{equation*}
In Fig.~\ref{F3}d, the mean IPR for each energy is shown.

\clearpage
\newpage

\begin{suppfigure}
    \centering
    \includegraphics[width=\textwidth,height=\textheight,keepaspectratio]{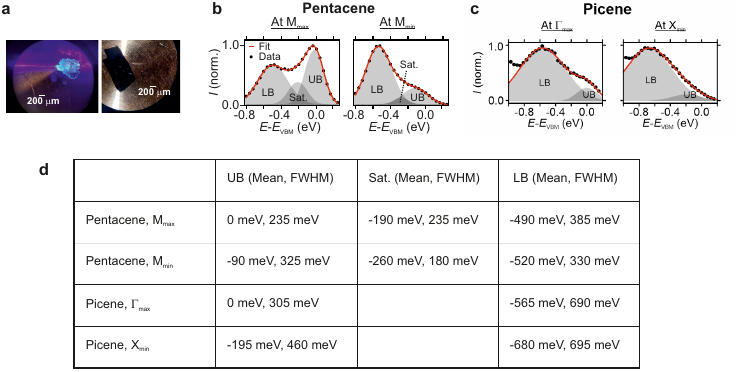}\\
    \caption{\textbf{\textbar~Crystals and energy distribution curves. a}, The picene (left) and pentacene crystals (right) used in the experiment. \textbf{b}, The energy distribution curves (EDC) for pentacene at maximum and minimum positions of the valence band. \textbf{c}, The EDCs for picene at the extremal positions of the valence band. \textbf{d}, The fit parameters of the EDCs.}
    \label{SuppFig:ARPES}
\end{suppfigure}

\begin{suppfigure}
    \centering
    \includegraphics[width=\textwidth,height=\textheight,keepaspectratio]{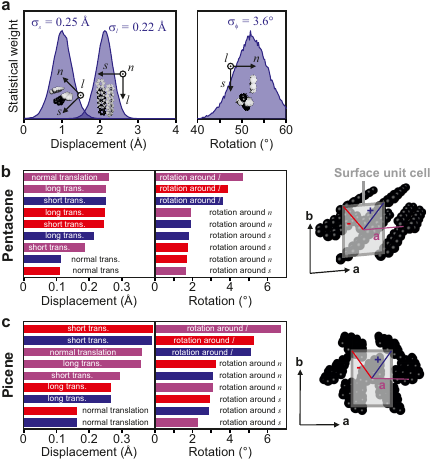}\\
    \caption{\textbf{\textbar~Molecular displacements from MD simulation. a},  Distribution of representative thermal displacements and rotations from a MD simulation at 295 K on a pentacene crystal. The coordinates refer to the local coordinate system of a molecular dimer of nearest neighbors along the (a+b)/2 direction. \textbf{b}, Extracted standard deviations $\sigma_r$ for displacements and rotations of all dimers (a, +, - with corresponding colors) in pentacene and \textbf{c} in picene.}
    \label{SuppFig:sigma_r}
\end{suppfigure}

\begin{suppfigure}
    \centering
    \includegraphics[width=\textwidth,height=\textheight,keepaspectratio]{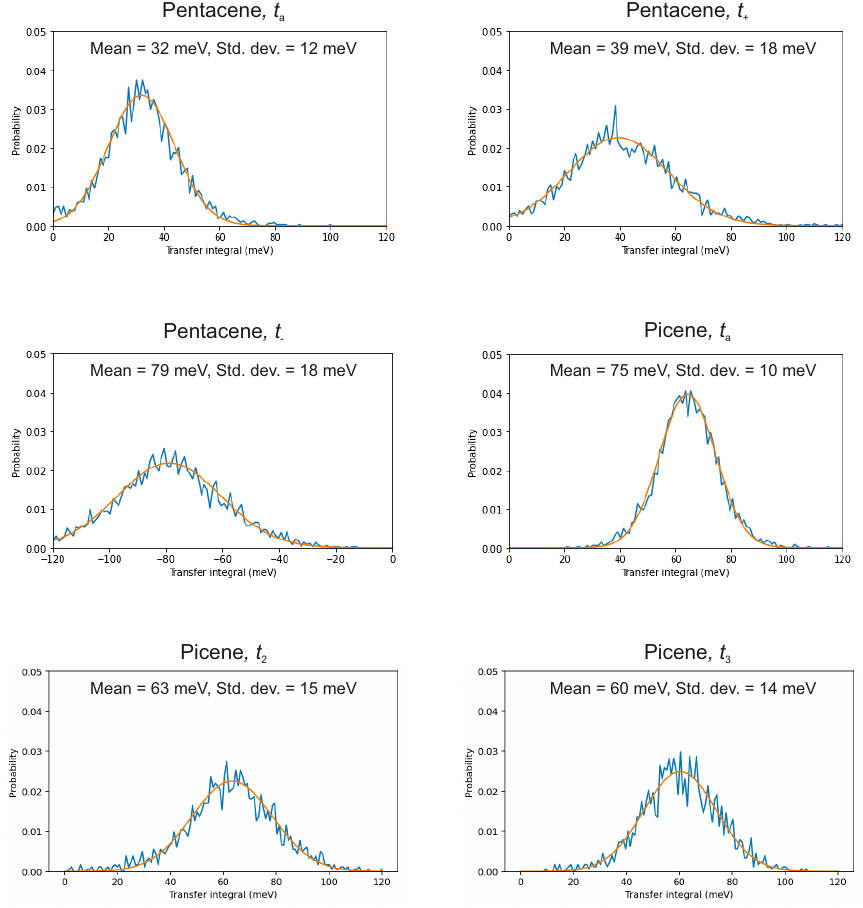}\\
    \caption{\textbf{\textbar~Statistics of the dimer transfer integrals from FO-DFT.} The dimer configurations were chosen from the MD simulations as detailed in the methods section.}
    \label{SuppFig:t_statistics}
\end{suppfigure}

\begin{suppfigure}
    \centering
    \includegraphics[width=\textwidth,height=\textheight,keepaspectratio]{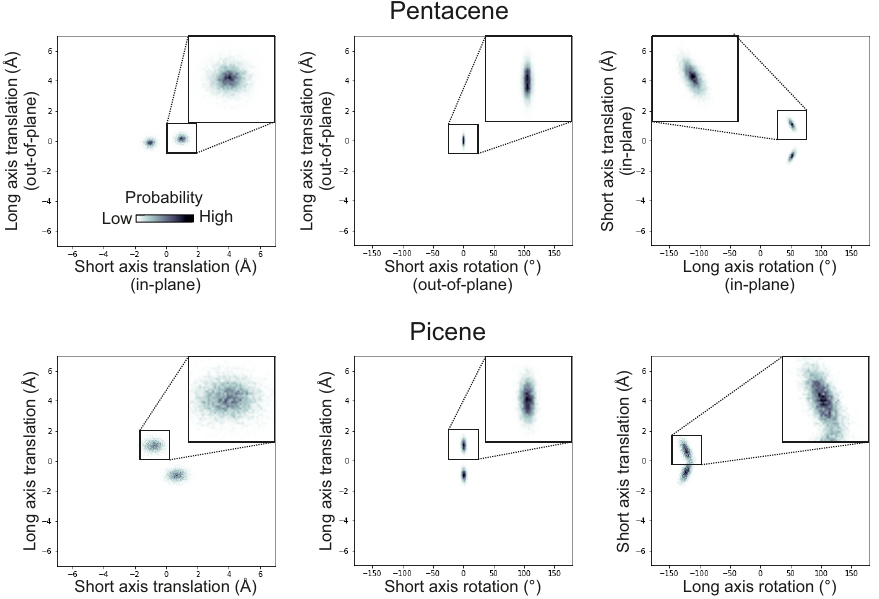}\\
    \caption{\textbf{\textbar~Correlation of DOF from MD simulation.} The panels show two-dimensional histograms of different DOF in pentacene and picene for the $(a+b)/2$-dimers. The first column shows that there is no correlation of out-of-plane and in-plane DOF. The second column shows a lack of correlation of the out-of-plane DOF. The in-plane DOF, however, are strongly correlated as the last column shows.}
    \label{SuppFig:MDcorrelation}
\end{suppfigure}

\begin{suppfigure}
    \centering
    \includegraphics[width=\textwidth,height=\textheight,keepaspectratio]{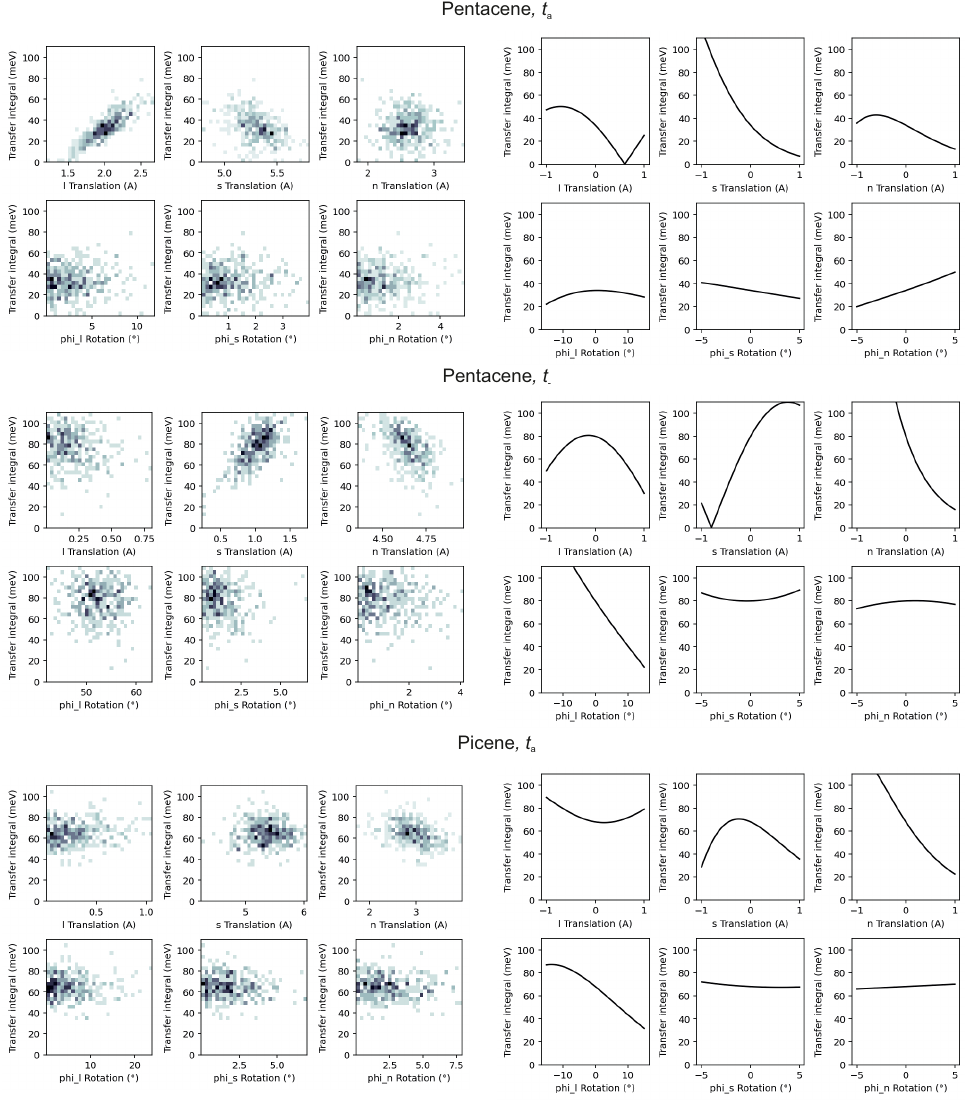}\\
    \caption{\textbf{\textbar~Correlation of the transfer integrals with the six DOF in selected dimers.} The figures to the left contain the dimer transfer integrals with the configurations given by the MD simulation, i.e. each transfer integral corresponds to a different coordinate $(l, s, n, \phi_l, \phi_s, \phi:n)$. The figures to the right contain the dependence of the transfer integrals on changing a single DOF while leaving the others unchanged. These curves are the basis for extracting the partial derivatives.}
    \label{SuppFig:t_r_correlation}
\end{suppfigure}

\clearpage
\newpage
\bibstyle{sn-standardnature}
\bibliography{sn-bibliography.bib}

\end{document}